\documentclass[twocolumn,prl,showpacs,superscriptaddress]{revtex4}
\usepackage{graphicx}
\usepackage{amsmath}
\usepackage{amssymb}

\begin{document}

\title{Stretching dependence of the vibration modes of a
single-molecule Pt-H$_2$-Pt bridge}
\author{D. Djukic}
\affiliation{Kamerlingh Onnes Laboratorium, Universiteit Leiden,
Postbus 9504, NL - 2300 RA Leiden, The Netherlands}
\author{K.~S. Thygesen}
\affiliation{Center for Atomic-scale Materials Physics,
  Department of Physics, Technical University of Denmark, DK - 2800
  Kgs. Lyngby, Denmark}
\author{C. Untiedt}
\altaffiliation[Present address: ]{Dpto. de F{\'\i}sica Aplicada,
Universidad de Alicante, E-03690 Alicante, Spain}
\affiliation{Kamerlingh Onnes Laboratorium, Universiteit Leiden,
Postbus 9504, NL - 2300 RA Leiden, The Netherlands}
\author{R.~H.~M. Smit}
\altaffiliation[Present address: ]{Dpto. de F{\'\i}sica de la
Materia Condensada - C3, Universidad Aut{\'o}noma de Madrid, 28049
Madrid, Spain } \affiliation{Kamerlingh Onnes Laboratorium,
Universiteit Leiden, Postbus 9504, NL - 2300 RA Leiden, The
Netherlands}
\author{K.~W. Jacobsen}
\affiliation{Center for Atomic-scale Materials Physics,
  Department of Physics, Technical University of Denmark, DK - 2800
  Kgs. Lyngby, Denmark}
\author{J.~M. van Ruitenbeek}
\email[Email: ] {ruitenbeek@physics.leidenuniv.nl}
\affiliation{Kamerlingh Onnes Laboratorium, Universiteit Leiden,
Postbus 9504, NL - 2300 RA Leiden, The Netherlands}
\date{\today}

\begin{abstract}
  A conducting bridge of a single hydrogen molecule between
  Pt electrodes is formed in a break junction experiment. It
  has a conductance near the quantum unit, $G_0=2e^2/h$,
  carried by a single channel. Using point contact spectroscopy
  three vibration modes   are observed and their variation upon
  isotope substitution is obtained. The stretching dependence
  for each of the modes allows uniquely classifying them as
  longitudinal or transversal modes. The interpretation of the
  experiment in terms of a $\text{Pt-H}_2\text{-Pt}$ bridge is
  verified by Density Functional Theory calculations for the
  stability, vibrational modes, and conductance of
  the structure.
\end{abstract}

\maketitle

There is beauty and power in the idea of constructing electronic
devices using individual organic molecules as active elements.
Although the concept was proposed as early as 1974~\cite{aviram74}
only recently experiments aimed at contacting individual organic
molecules are being reported
\cite{reed97,kergueris99,reichert02,park02,cui02,bumm96,%
liang02,kubatkin03,xu03,kervennicUNPUB} and devices are being
tested \cite{luo02xx,collier99}. The first results raised high
expectations, but quickly problems showed up such as large
discrepancies between the current-voltage characteristics obtained
by different experimental groups, and large discrepancies between
experiments and theory. The main tools that have been applied in
contacting single molecules are STM (or conducting tip AFM) and
break junction devices. Often it is difficult to show that the
characteristics are due to the presence of a molecule, or that
only a single molecule has been contacted. There has been
important progress in analysis and reproducibility of some
experiments~\cite{reichert02,park02,liang02,kubatkin03,xu03,kervennicUNPUB},
but in comparing the data with theory many uncertainties remain
regarding the configuration of the organic molecule and the nature
of the molecule-metal interface. The organic molecules selected
for these studies are usually composed of several carbo-hydride
rings and are anchored to gold metal leads by sulphur groups. In
view of the difficulties connected with these larger molecules it
seems natural to step back and focus on even simpler systems.

Here we concentrate on the simplest molecule, $\text{H}_2$,
anchored between platinum metal leads using mechanically
controllable break junctions. The first experiments on this
system~\cite{smit02} showed that the conductance of a single
hydrogen molecule between Pt leads is slightly below $1G_0$, where
$G_0=2e^2/h$ is the conductance quantum. A vibration mode near
65~meV was observed and interpreted as the longitudinal
center-of-mass (CM) mode of the molecule. These results have
inspired new calculations on this problem using Density Functional
Theory (DFT) methods \cite{cuevas03,garcia04}. Cuevas {\it et
al.}~\cite{cuevas03} find a conductance around $0.9G_0$, in
agreement with the first DFT calculations presented
in~\cite{smit02}. In contrast, Garc{\'\i}a {\it et
al.}~\cite{garcia04} obtain a conductance of only $(0.2-0.5)~G_0$
for the in-line configuration of the hydrogen molecule. Instead
they propose an alternative configuration with hydrogen atoms
sitting above and below a Pt-Pt atomic contact.

In this letter we combine new experimental results with DFT
calculations to show that the configuration proposed
in~\cite{smit02} is correct, yet the observed vibration mode was
incorrectly attributed. In contrast, the present experiment
resolves three vibration modes that can be classified as
longitudinal or transverse modes based on the observed shifts with
stretching of the contacts. The comparison with the calculations
is nearly quantitative and the large number of experimentally
observed parameters (the number of vibration modes, their
stretching dependence and isotope shifts, the conductance and the
number of conductance channels) puts stringent constraints on any
possible interpretation.

The measurements have been performed using the mechanically
controllable break junction technique~\cite{muller92a,agrait03}. A
small notch is cut at the middle of a Pt wire to fix the breaking
point. The wires used are $100~\mu$m in diameter, about 1~cm long,
and have a purity of 99,9999\%. The wire is glued on top of a
bending beam and mounted in a three-point bending configuration
inside a vacuum chamber. Once under vacuum and cooled to 4.2~K the
wire is broken by mechanical bending of the substrate. Clean
fracture surfaces are exposed and remain clean for days in the
cryogenic vacuum. The bending can be relaxed to form atomic-sized
contacts between the wire ends using a piezo element for fine
adjustment.

After admitting a small amount ($\sim 3~\mu$mole) of molecular
H$_{2}$ (99.999\%) in the sample chamber and waiting some time for
the gas to diffuse to the cold end of the insert, a sudden change
is observed in the conductance of the last contact before
breaking. The typical value of $(1.6 \pm 0.4)G_{0}$ for a
single-atom Pt contact is replaced by a frequently observed
plateau near $1G_{0}$ that has been attributed in~\cite{smit02} to
the formation of a Pt-H$_{2}$-Pt bridge. By increasing the bias
voltage above 300~mV we recover the pure Pt conductance. But as
soon as the bias voltage is decreased the H$_{2}$ induced plateaus
at $1G_{0}$ reappear. We interpret this as desorption of hydrogen
due to Joule heating of the contacts. For biases below 100~mV, the
Pt-H$_{2}$-Pt bridge can be stable for hours.

\begin{figure}[!t]
\includegraphics[width=\linewidth]{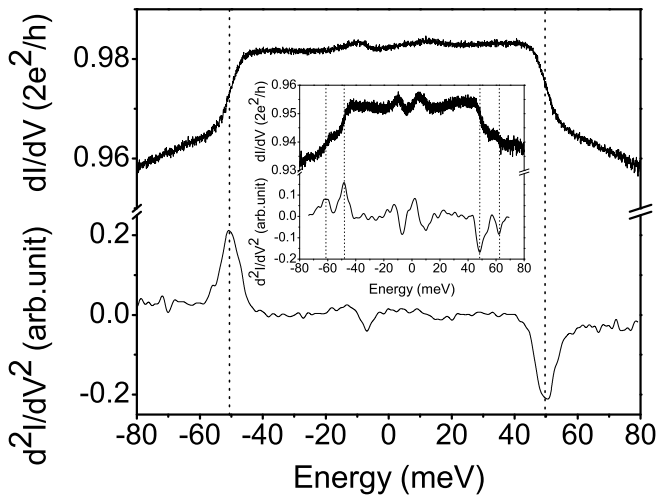}
\caption{Differential conductance curve for D$_{2}$ contacted by
Pt leads. The $dI/dV$ curve (top) was recorded over 1 minute,
using a standard lock-in technique with a voltage bias modulation
of 1~meV at a frequency of 700Hz. The lower curve shows the
numerically obtained derivative. The spectrum for H$_{2}$ in the
inset shows two phonon energies, at 48 and 62~meV. All spectra
show some, usually weak, anomalies near zero bias that can be
partly due to excitation of modes in the Pt leads, partly due to
two-level systems near the contact \cite{kozub86}.
\label{fig.D2spectrum}}
\end{figure}

At the $1G_0$-conductance plateaus we take differential
conductance ($dI/dV$) spectra in order to determine the inelastic
scattering energies. By repeatedly breaking the contacts, joining
them again to a large contact, and pulling until arriving at a
plateau near $1G_0$, we obtain a large data set for many
independent contacts. The experiments were repeated for more than
5 independent experimental runs, and for the isotopes HD (96\%)
and D$_{2}$ (99.7\%). Fig.~\ref{fig.D2spectrum} shows a spectrum
taken for D$_2$ showing a sharp drop in the differential
conductance by 1--2\% symmetrically at $\pm 50$~meV. Such signals
are characteristic for point contact
spectroscopy~\cite{khotkevich95}, which was first applied to
single-atom contacts by~\cite{agrait02a}. The principle of this
spectroscopy is simple: when the difference in chemical potential
between left- and right-moving electrons, $eV$, exceeds the energy
of a vibration mode, $\hbar\omega$, back-scattering associated
with the emission of a vibration becomes possible, giving rise to
a drop in the conductance. This can be seen as a dip (peak) in the
second derivative $d^2I/dV^2$ at positive (negative) voltages, as
in Fig.~\ref{fig.D2spectrum}.

\begin{figure}[!b]
\includegraphics[width=0.8\linewidth]{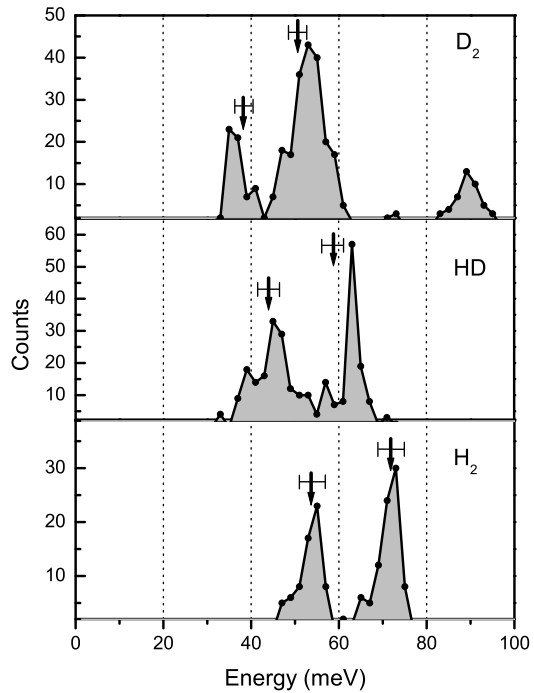}
\caption{Distribution of vibration mode energies observed for
H$_{2}$, HD, and D$_{2}$ between Pt electrodes, with a bin size of
2~meV. The peaks in the distribution for H$_{2}$ are marked by
arrows and their widths by error margins. These positions and
widths were scaled by the expected isotope shifts, $\sqrt{2/3}$
for HD and $\sqrt{1/2}$ for D$_2$, from which the arrows and
margins in the upper two panels have been obtained.
\label{fig.frequency-histo}}
\end{figure}

Some contacts can be stretched over a considerable distance, in
which case we observe an {\em increase} of the vibration mode
energy with stretching. This observation suffices to invalidate
the original interpretation~\cite{smit02} of this mode as the
longitudinal CM mode. Indeed, our DFT calculations show that the
stretching mainly affects the Pt-H bond which is elongated and
weakened resulting in a drop in the frequency of the $\text{H}_2$
longitudinal CM mode. An increase can be obtained only for a
transverse mode which, like a guitar string, obtains a higher
pitch at higher string tension due to the increased restoring
force.

On many occasions we observe two modes in the $dI/dV$ spectra, see
the inset of Fig.~\ref{fig.D2spectrum}. The relative amplitude of
the two modes varies; some spectra show only the lower mode, some
only the higher one. All frequencies observed in a large number of
experiments are collected in the histograms shown in
Fig.~\ref{fig.frequency-histo}. With a much larger data set
compared to~\cite{smit02} we are now able to resolve two peaks in
the distribution for H$_2$ corresponding to the two modes seen in
the inset of Fig.~\ref{fig.D2spectrum}. The peaks are expected to
shift with the mass $m$ of the isotopes as
$\omega\propto\sqrt{1/m}$. This agrees with the observations, as
shown by the scaled position of the hydrogen peaks marked by
arrows above the distributions for D$_2$ and HD. Note that the
distribution for HD proves that the vibration modes belong to a
molecule and not an atom, since the latter would have produced a
mixture of the H$_2$ and D$_2$ distributions.
\begin{figure}[!b]
\includegraphics[width=0.9\linewidth]{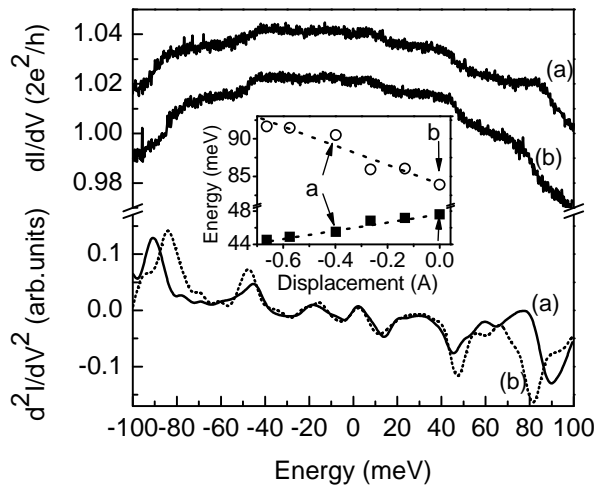}
\caption{$dI/dV$ Spectrum (top) for Pt-D$_2$ junctions, and their
numerical derivatives $d^2I/dV^2$ (bottom). The spectrum (b) was
obtained on the same contact as (a) after stretching the junction
by 0.04~nm. They are two spectra out of a series of six and the
complete evolution of the two modes is shown in the inset.
\label{fig.stretching-dependence} }
\end{figure}
In the case of D$_2$ we observe a third peak in the distribution
at 86--92~meV. For the other isotopes this mode falls outside our
experimentally accessible window of about $\pm100$~meV, above
which the contacts are destabilized by the large current.
Fig.~\ref{fig.stretching-dependence} shows the dependence of this
mode upon stretching of the junction. In contrast to the two
low-frequency modes this mode shifts down with stretching,
suggesting that this could be the longitudinal CM mode that was
previously attributed to the low-frequency modes~\cite{smit02}.

\begin{figure}[!t]
\includegraphics[width=0.8\linewidth]{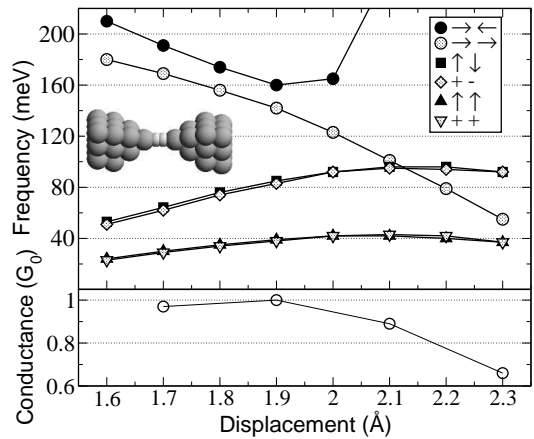}
\caption{Calculated vibrational frequencies for the hydrogen atoms
in the contact as a function of elongation of the supercell. The
inset shows the atomic arrangement in the supercell. The lower panel
shows the calculated conductance.
\label{fig.calculated-frequencies} }
\end{figure}

In order to test the interpretation of the experiment in terms of
a Pt-$\text{H}_2$-Pt bridge we have performed extensive DFT
calculations using the plane wave based pseudopotential code
Dacapo~\cite{bahn02xx,hammer99}. The molecular contact is
described in a supercell containing the hydrogen atoms and two
4-atom Pt pyramids attached to a Pt(111) slab containing four
atomic layers, see inset of Fig.~\ref{fig.calculated-frequencies}
\cite{compdetails}. In the total energy calculations both the
hydrogen atoms and the Pt pyramids were relaxed while the
remaining Pt atoms were held fixed in the bulk structure. The
vibration frequencies were obtained by diagonalizing the Hessian
matrix for the two hydrogen atoms. The Hessian matrix is defined
by $\partial^2 E_0/(\partial \tilde u_{n\alpha}
\partial \tilde u_{m\beta} )$, where $E_0$ is the ground state
potential energy surface and $\tilde u_{n\alpha}$ is the
displacement of atom $n$ in direction $\alpha$ multiplied by the
mass factor $\sqrt{M_n}$. In calculating the vibration modes all
Pt atoms were kept fixed which is justified by the large
difference in mass between H and Pt. The conductance is calculated
from the Meir-Wingreen formula~\cite{meirwingreen} using a basis
of partly occupied Wannier functions~\cite{powannier},
representing the leads as bulk Pt crystals.

In order to simulate the stretching process of the experiment we
have calculated contacts for various lengths of the supercell. The
bridge configuration is stable over a large distance range with
the binding energy of the $\text{H}_2$ molecule varying from
$-0.92$~eV to $-0.47$~eV, relative to gas phase $\text{H}_2$ and a
broken Pt contact, over the range of stretching considered here.
The H-H bond length stays close to 1.0~{\AA} during the first
stages of the stretching upon which it retracts and approaches the
value of the free molecule. The hydrogen thus retains its
molecular form and the elongation mainly affects the weaker Pt-H
bond. For smaller electrode separations a structure with two
hydrogen atoms adsorbed on the side of a Pt-Pt atomic contact
becomes the preferred geometry, as also found by Garc{\'\i}a {\it
et al. }~\cite{garcia04}. However, we find that the latter
structure has a conductance of 1.5~$G_0$, well above 1~$G_0$.
Moreover, this structure has at least three conduction channels
with significant transmission, which excludes it as a candidate
structure based on the analysis of conductance fluctuations in
\cite{smit02} and \cite{csonka04}, which find a single channel
only. In view of the activity of the Pt surface towards catalyzing
hydrogen dissociation one would have anticipated a preference for
junctions of hydrogen in atomic form. However, we find that the
bonding energy of H compared to that of H$_2$ strongly depends on
the metal coordination number of the Pt atom. For metal
coordination numbers smaller than 7 bonding to molecular hydrogen
is favored, the bond being strongest for fivefold coordinated Pt.

The calculations identify the six vibrational modes of the
hydrogen molecule.  For moderate stretching two pairwise
degenerate modes are lowest in frequency. The lowest one
corresponds to translation of the molecule transverse to the
transport direction while the other one is a hindered rotation
mode. The two modes are characterized by increasing frequencies as
a function of stretch of the contact. At higher energies we find
the two longitudinal modes: first the CM mode and then the
internal vibration of the molecule. These two modes become softer
during stretching up to Pt-H bond lengths of about 1.9~{\AA}.
Beyond this point the Pt-H bond begins to break and the internal
vibration mode approaches the one of the free molecule.

The variation of the frequencies of the lowest lying hydrogen
modes with stretching is thus in qualitative agreement with the
experiments, a strong indication that the suggested structure is
indeed correct. The agreement is even semiquantitative: If we
focus on displacements in the range 1.7 - 2.0 {\AA} (see Fig.~4)
the calculated conductance does not deviate significantly from the
experimentally determined value close to $1G_0$.  In this regime
the three lowest calculated frequencies are in the range 30-42
meV, 64-92 meV, and 123-169 meV. The two lowest modes can be
directly compared with the experimentally determined peaks at 54
and 72 meV observed for H$_2$, while a mass re-scaling of the
D$_2$ result for the highest mode gives approximately 126 meV.

The second peak in the HD distribution in
Fig.~\ref{fig.frequency-histo} is somewhat above the position
obtained by scaling the $\text{H}_2$ peak by $\sqrt{3/2}$. The
transverse translation mode and the hindered rotation mode are
decoupled when the two atoms of the molecule have the same mass.
In the case of HD they couple and the simple factor does not hold.
Having identified the character of these modes a proper re-scaling
of the experimentally determined H$_2$ frequencies (54 and 72 meV)
to the case of HD leads to the frequencies 42 and 66 meV, in very
good agreement with the peaks observed for HD.

Even though there is good agreement with the calculated signs of
the frequency shifts with stretching for the various modes, there
is a clear discrepancy in magnitudes. Considering, e.g., the
high-lying mode for D$_2$, the measured shift of the mode is of
the order 15 meV/{\AA}, which is almost an order of magnitude smaller
than the calculated variation of around 130 meV/{\AA}. However,
experimentally the distances are controlled quite far from the
molecular junction and the elastic response of the electrode
regions have to be taken into account. Simulations of atomic chain
formation in gold \cite{chainformation} during contact breaking
show that most of the deformation happens not in the atomic chains
but in the nearby electrodes. A similar effect for the Pt-H$_2$-Pt
system will significantly reduce the stretch of the molecular
bridge compared to the displacement of the macroscopic electrodes.

The observation of the three vibration modes and their stretching
dependence provides a solid basis for the interpretation. The
fourth  mode, the internal vibration, could possibly be observed
using the isotope tritium. The hydrogen molecule junction can
serve as a benchmark system for molecular electronics
calculations. The experiments should be gradually expanded towards
more complicated systems and we have already obtained preliminary
results for CO and C$_2$H$_2$ between Pt leads.

We thank M. Suty for assistance in the experiments and M. van
Hemert for many informative discussions. This work was supported
by the Dutch ``Stichting FOM,'', the Danish Center for Scientific
Computing through Grant No. HDW-1101-05 and the Spanish MCyT under
contract MAT-2003-08109-C02-01 and the ESF through the EUROCORES
SONS programme.

\bibliographystyle{apsrev}

\begin{thebibliography}{26}
\bibitem{compdetails} The plane wave expansion is cut off at 25 Ry. We use ultrasoft
  pseudopotentials~\cite{vanderbilt90xx} and the exchange correlation
  functional PW91~\cite{perdew92}. A (1,4,4) Monkhorst pack grid has
  been used to sample the Brillouin zone.
\end{thebibliography}

\end{document}